\documentclass[%
 reprint,
amsmath,
amssymb,
pra,
floatfix,
balancelastpage,
]{revtex4-1}
\usepackage{color}
\usepackage{epstopdf}
\usepackage{graphicx}
\usepackage{dcolumn}
\usepackage{bm}
\usepackage{amssymb}   
\usepackage{amsmath}
\usepackage{braket}
\usepackage{csquotes}
\usepackage{hyperref}

\begin{document}

\preprint{APS/123-QED}

\title{Entanglement dynamics of two coupled mechanical oscillators in modulated optomechanics}

\author{Subhadeep Chakraborty}
 \email{c.subhadeep@iitg.ernet.in}
\author{Amarendra K. Sarma}%
 \email{aksarma@iitg.ernet.in}
\affiliation{%
 Department of Physics, Indian Institute of Technology  Guwahati, Guwahati-781039, Assam, India}


%

\date{\today}

\begin{abstract}

We study the entanglement dynamics of two coupled mechanical oscillators, within a modulated optomechanical system. We find that, depending on the strength of the mechanical coupling, one could observe either a stationary or a dynamical behavior of the mechanical entanglement, which is extremely robust against the oscillator temperature. Moreover, we have shown that this entanglement dynamics is strongly related to the stability of the normal modes. Taking mechanical damping effects into account, an analytical expression corresponding to the critical mechanical coupling strength, where the transition from stationary to dynamical entanglement occurs is also reported. The proposed scheme is analysed with experimentally realistic parameters, making it a promising mean to realize macroscopic quantum entanglement within current state-of-the-art experimental setups.
\begin{description}
\item[PACS numbers]
\verb|42.50.Pq,42.50.Lc,42.50.Dv,03.67.Bg|
\end{description}
\end{abstract}

\pacs{Valid PACS appear here}
\maketitle


\section{\label{intro}INTRODUCTION}

The generation of quantum entanglement between two macroscopic, massive objects has been a task of paramount importance, both in fundamental research, in particular for fundamental tests of quantum mechanics \cite {leggett, zurek, bassi, zhang} and in numerous futuristic potential applications related to quantum computing, quantum information processing, quantum communication and so on \cite{rabl, conti, man_tom, xureb, chang}. Thanks to the fast growing field of cavity optomechanics \cite{aspelmeyer, meystre}, which provides a versatile platform to prepare such an entangled state in mechanical motions, research in the area has virtually exploded in recent times. Relying on the generic radiation-pressure coupling, much studies have already been reported on the entanglement generation between a cavity field and a mechanical oscillator \cite{vitali, genes, abdi, palomaki, kuzyk, hofer, wang, flipp}. Besides, based on similar architecture, a lot of emphasis has been currently brought forward to realize entanglement between two macroscopic mechanical oscillators. These studies mostly include: light-to-matter entanglement transfer \cite{mancini, sete}, driving the optical cavity with a two-tone field \cite{li, li_2}, dissipation induced optomechanical entanglement \cite{tan, r_x_chen} and the reservoir engineered based schemes \cite{r_yang, r_wang, woolley, asjad, yan}. However, in most of the cases, the possibility of observing such a nonclassical state is seriously hindered by the presence of the environmental noise. Hence a lot of emphasis has currently been put in realizing quantum entanglement at higher bath temperature. 

While the search for robust and \textit{hot entanglement} \cite{hot} in optomechanical systems is on, it occurs that modulating an optomechanical system may be a very rewarding proposition in achieving a more robust nonclassicality. For example, in 2009, Mari \textit{et al.} \cite{mari} first showed that by gently modulating an optomechanical system, one could not only enhance the degree of squeezing in mechanical quadratures but also improve the stationary entanglement between the cavity field and the nanomechanical resonator. Following Ref. \cite{mari}, Farace \textit{ et al.} \cite{farace} studied the effect of both the amplitude modulation and frequency modulation in an optomechanical system. They showed that there exists an optimal modulation regime where the desired quantum effects can either be enhanced or suppressed. Along this line, Schmidt \textit{et al.} \cite{schmidt} implemented a suitable amplitude modulation scheme in optomechanical circuits for continuous variable (CV) quantum state processing. More recently, Chen \textit{ et al.} \cite{chen} has exploited the same amplitude modulation to improve the stationary mechanical entanglement in a double cavity optomechanical system.

In parallel to the developments in optomechanical systems, one modulation scheme of particular interest is the so-called periodic modulation of the coupling strength. It is now well established that by periodically driving the coupling strength with a frequency twice that of the oscillator frequency, one can squeeze the collective quadratures, leading to entanglement generation between the two harmonic oscillators.  In 2010, within a similar framework, two identical harmonic oscillators in contact with two independent thermal baths and coupled via a time periodic driving, Galve \textit{ et al.} \cite{galve} first demonstrated the existence of stationary entanglement at a relatively high temperature. Following Ref. \cite{galve}, Roque\textit{ et al.} \cite{roque} reported the dynamics of quantum correlations between two coupled harmonic oscillators in contact with a common heat bath. They found that it is not the bath temperature, rather the system parameters to which the entanglement dynamics is more sensitive. However, it should be noted that in their study they could not find any steady-state behavior of the generated entanglement. On the other hand, recently, Chen \textit{ et al.} \cite{chen_ent_har} considered a system of two coupled harmonic oscillators connected via a weak time-dependent coupling. In the absence of any environmental decoherence, they reported that a transition from bounded to unbounded entanglement dynamics occurs when the modulation strength crosses a critical value.

In this work, we theoretically study the entanglement dynamics of two coupled mechanical oscillators, placed within a modulated optomechanical system.  We show that, unlike Ref. \cite{roque}, both the stationary and dynamical behavior of the mechanical entanglement could be achieved. Moreover, by taking the mechanical damping terms into account, we give an analytical estimation of the critical mechanical coupling strength where the dynamical transition occurs. The rest of the paper is organized as follows. In Sec. \ref{model} we introduce our physical model of the optomechanical system and derive the equation of motion corresponding to the correlation matrix. In Sec. \ref{results} we give a detailed discussion of the dynamical behavior of the mechanical entanglement for various mechanical coupling strengths, and, show the connection between the entanglement dynamics and the stability of the normal modes. Finally, we present our concluding remarks in Sec. \ref{conclusion}.

\begin {figure}[t]\label{sys}
\begin {center}
\includegraphics [width =7cm]{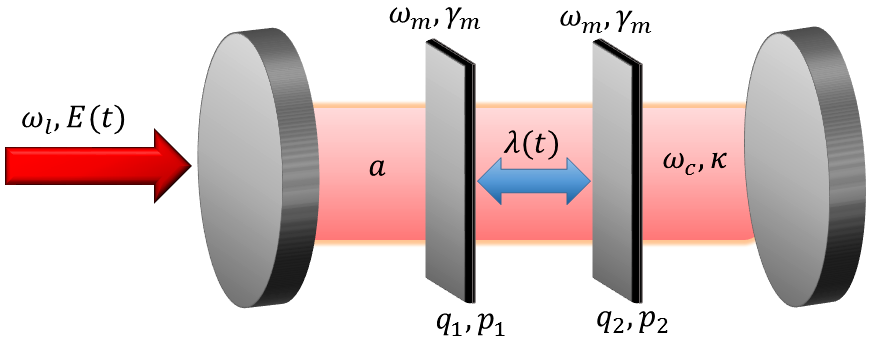}
\caption{\label {sys}(Color online) Schematic diagram of the considered optomechanical system. The cavity is driven by an amplitude modulated laser $E(t)$, and, the two mechanical oscillators are coupled by a time-periodic driving $\lambda(t)$.}
\end{center}
\end{figure}
\section{\label{model}MODEL AND DYNAMICS}

The system under consideration, consists of two identical mechanical oscillators placed within an optical cavity. An external laser with a time-dependent amplitude $E(t)$ and frequency $\omega_l$ drives the cavity which makes the two oscillators couple indirectly via the radiation-pressure interaction. Besides, there is a direct mechanical coupling between the two oscillators, with a periodically modulated coupling strength $\lambda(t)$. The schematic of our system is depicted in Fig. 1, and, the Hamiltonian (in a frame rotating with the laser frequency) is written as follows (in the unit of $\hbar=1$): 
\begin{gather}\label{hamiltonian}
 H=\Delta_0 a^\dagger a+\sum_{j=1}^{2}\frac{\omega_m}{2}(q_j^2+p_j^2)+g a^\dagger a \sum_{j=1}^2q_j+\lambda(t)q_1q_2\\\nonumber
 +i\left(E(t)a^\dagger-E^*(t)a\right).
\end{gather}
Here, $a$ ($a^\dagger$) refers to the annihilation (creation)  operator of the cavity field (with frequency $\omega_c$ and decay rate $\kappa$), $q_j$ ($p_j$) is the dimensionless position (momentum)  operator of the $j$-th mechanical oscillator (with frequency $\omega_m$ and damping rate $\gamma_m$).  $g$ refers to the strength of the single photon radiation-pressure coupling. In Hamiltonian \eqref{hamiltonian}, the first and the second term corresponds to the Hamiltonian of the driven cavity and the mechanical oscillators, respectively, with $\Delta_0=\omega_c-\omega_l$ being the optical detuning. The third term describes the optomechanical interaction between the cavity field and the mechanical oscillators, while the fourth term refers to the bilinear coupling between the two oscillators. Finally, the last term gives the external laser driving.

In addition to this, the system dynamics is unavoidably subjected to the fluctuation-dissipation processes affecting both the cavity field and the mechanical oscillators. Taking all the damping and noise terms into account, the dynamics of the system is fully described by the following set of nonlinear quantum Langevin equations (QLEs):
\begin{subequations}\label{langevin}
\begin{align}
\dot{q_j} &=\omega_m p_j,\\
\dot{p_j} &=-\omega_m q_j-g a^\dagger a -\lambda(t) q_{3-j}-\gamma_m p_j+\xi_j(t),\\
\dot{a} &=-\{i(\Delta_0+g\sum_{j=1}^2q_j)+\kappa\}a+E(t)+\sqrt{2\kappa}a_{in},
\end{align}
\end{subequations}
where $a_{in}$ is the vacuum input noise operator, with the only nonzero correlation function \cite{a_in}:
\begin{equation}
 \langle a_{in}(t)a^\dagger _{in}(t^\prime)\rangle=\delta(t-t^\prime).
\end{equation}
 $\xi_j(t)$ are the stochastic Hermitian Brownian noise operator, with the non-Markovian correlation function given by \cite{brownian}
\begin{gather}
 \langle \xi_j(t)\xi_k(t^\prime)\rangle=\\\nonumber
 \frac{\delta_{jk}}{2\pi}\frac{\gamma_m}{\omega_m}\int \omega e^{-i\omega(t-t^\prime)}\left[\coth\left(\frac{\hbar \omega}{2K_BT}\right)+1\right]d\omega,
\end{gather}
($k_B$ being the Boltzmann constant and $T$ being the temperature of the mechanical oscillators). However, in the limit of large mechanical quality factor $Q=\omega_m/\gamma_m\gg1$, one could well approximate this Brownian noise to a Markovian delta-correlated relation \cite{brownian_delta}:
\begin{equation}
\langle\xi_j(t)\xi_j(t^\prime)+\xi_j(t^\prime)\xi_j(t)\rangle /2\simeq\gamma_m\left(2n_{th}+1\right)\delta\left(t-t^\prime\right),
\end{equation}
with $n_{th}=\left[\mathrm{ exp}\left(\frac{\hbar \omega_m}{K_BT}\right)-1\right]^{-1}$ being the number of mean thermal phonons.

Next, when the system is strongly driven to a large classical mean value, we can adopt the standard linearization technique and rewrite each Heisenberg operator as follows: $o(t)=\langle o(t) \rangle+\delta o(t)$ ($o=q_j, p_j, a$). Here, $\langle o(t)\rangle$ refers to the classical $c$-number mean value and $\delta o(t)$ is the zero-mean quantum fluctuation around the classical mean value. The equation of motion corresponding to the classical mean values is given by the following set of nonlinear differential equations:
\begin{subequations}\label{steady}
\begin{align}
\langle \dot{q_j}(t) \rangle & =\omega_m \langle p_j(t)\rangle,\\
\langle \dot{p_j}(t) \rangle & =-\omega_m \langle q_j(t) \rangle-g|\langle a(t) \rangle|^2-\lambda(t) \langle q_{3-j}(t)\rangle \\\nonumber & -\gamma_m \langle p_j(t) \rangle,\\
\langle \dot{a}(t) \rangle &=-\{i(\Delta_0+g\sum_{j=1}^2 \langle q_j(t) \rangle)+\kappa\}\langle a(t) \rangle +E(t).
\end{align}
\end{subequations}
On the other hand, the dynamics of the quantum fluctuations is governed by the following linearized QLEs, written in a matrix form:
\begin{equation}\label{fluc}
 \dot{u(t)}=A(t)u(t)+n(t).
\end{equation}
Here, $u^T(t)=\left(\delta q_1(t),\delta p_1(t),\delta q_2(t),\delta p_2(t),\delta X(t),\delta Y(t)\right)$ is the vector of quadrature fluctuation operators,
$n^T(t)=\left(0,\xi_1(t),0,\xi_2(t),\sqrt{2\kappa}X_{in}(t),\sqrt{2\kappa}Y_{in}(t)\right)$ is the vector of corresponding noises and
\begin{align}\label{drift}
 A(t)=\left(
 \begin{array}{cccccc}
 0 & \omega_m & 0 & 0 & 0 & 0\\
 -\omega_m & -\gamma_m & -\lambda(t) & 0 & G_x(t) & G_y(t)\\
 0 & 0 & 0 & \omega_m & 0 & 0\\
 -\lambda(t) & 0 & -\omega_m & -\gamma_m & G_x(t) & G_y(t)\\
 -G_y(t) & 0 & -G_y(t) & 0 & -\kappa & \Delta(t)\\
 G_x(t) & 0 & G_x(t) & 0 & -\Delta(t) & -\kappa\\
 \end{array}
 \right),
\end{align}
is the drift matrix. The time-dependent coupling and the detuning terms, respectively, are defined as follows:
\begin{subequations}
\begin{align}
G(t) &=-\sqrt{2}g\langle a(t) \rangle,\\
G(t) &=G_x(t)+iG_y(t),\\ 	
\Delta(t) &=\Delta_0+g\sum_{j=1}^2\langle q_j(t)\rangle,
\end{align}
\end{subequations}
It should be noted that in Eq. \eqref{fluc} we have used the quadrature operators for the cavity field with the corresponding Hermitian input noise operators, respectively defined as:
$\delta X\equiv\frac{\left(\delta a+\delta a^\dagger\right)}{\sqrt2}$, $\delta Y\equiv\frac{\left(\delta a-\delta a^\dagger\right)}{i\sqrt2}$, and
$X_{in}\equiv\frac{\left(a_{in}+a_{in}^\dagger\right)}{\sqrt2}$, $Y_{in}\equiv\frac{\left(a_{in}-a_{in}^\dagger\right)}{i\sqrt2}$.

Due to the above linearized dynamics and the zero-mean Gaussian nature of the quantum noises, the quantum fluctuations in the stable regime evolve to an asymptotic Gaussian state which is completely characterized by its $6\times6$ correlation matrix, given by:
\begin{equation}\label{corr}
V_{ij}=\left(\langle u_i(t)u_j(t)+u_j(t)u_i(t)\rangle\right)/2.
\end{equation}
The equation of motion corresponding to the correlation matrix, using Eq. \eqref{fluc} and Eq. \eqref{corr}, can be written as follows:
\begin{equation}\label{eqm_corr}
 \dot{V}(t)=A(t)V(t)+V(t)A^T(t)+D,
\end{equation}
where $D=\mathrm{Diag}\left[0,\gamma_m(2n_{th}+1),0,\gamma_m(2n_{th}+1),\kappa,\kappa\right]$ is the matrix of noise-correlation. Note that, Eq. \eqref{eqm_corr} is an inhomogeneous first-order differential equation with 21 elements which could be numerically solved with the initial condition $V(0)=\mathrm{Diag}\left[n_{th}+1/2,n_{th}+1/2,n_{th}+1/2,n_{th}+1/2,1/2,1/2\right]$. Here, we have assumed that each mechanical oscillators are prepared in their thermal states at temperature $T$ and the cavity field is in it's vacuum state.
\begin {figure}[t]\label{stable}
\begin {center}
\includegraphics [width =7cm]{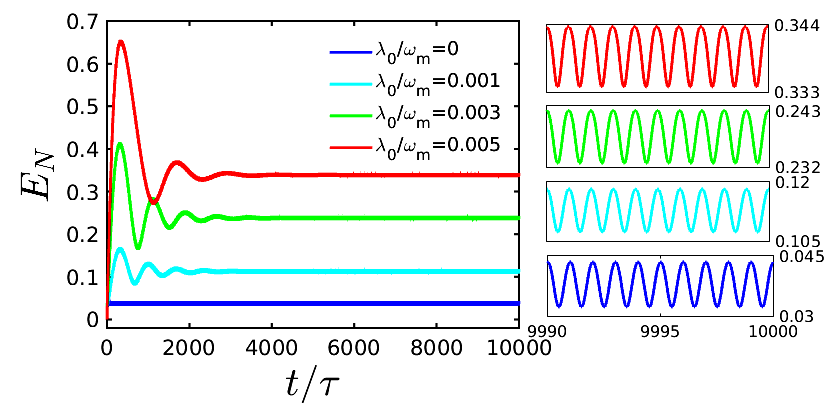}
\caption{\label {en_stable}(Color online) Entanglement dynamics of the two coupled mechanical oscillators for $\lambda_0/\omega_m=[0.001,0.005]$ at $T=0$. The left hand panel shows the asymptotic nature of the mechanical entanglement}
\end{center}
\end{figure}
\section{\label{results} RESULTS AND DISCUSSIONS}

Before proceeding to a direct numerical investigation of the mechanical entanglement, we first specify the exact form of time modulation for both the external driving and the mechanical coupling strength, as follows:
\begin{equation}
E(t)=E_0+E_1\mathrm{cos}\left(\Omega t\right), \quad \lambda(t)=\lambda_0\mathrm{cos}\left(\Omega t\right).
\end{equation}
Moreover, we choose the following set of parameters for our numerical simulations: $\Delta_0/\omega_m=1.0$, $\kappa/\omega_m=0.1$, $\gamma_m/\omega_m=5\times10^{-4}$, $g/\omega_m=1\times10^{-5}$, $E_0/\omega_m=1\times10^4$, $E_1/\omega_m=1\times10^3$, $T_0/\omega_m=\hbar/k_B$, $\Omega/\omega_m=2.003$ (this particular choice will be justified later) and $\tau=2\pi/\Omega$.

In Fig. 2, we plot the time evolution of the mechanical entanglement $E_N$ (see appendix \ref{gaussian}) for multiple values of $\lambda_0/\omega_m$. It can be observed that in absence of the mechanical coupling ($\lambda_0/\omega_m=0$), the two oscillators exhibits a very small degree of stationary entanglement. However, as soon as the mechanical coupling is introduced, there is a sudden but significant enhancement in $E_N$ at initial time, which finally converges to an asymptotic steady-state value. We note that this enhancement becomes more profound with an increase in coupling strength $\lambda_0/\omega_m$. Moreover, in the asymptotic regime, we find that the entanglement acquires the same period of modulation (see the right hand panel of Fig. 2). Hence one can identify the degree of the entanglement as the maximum over one period $\tau=2\pi/\Omega$ of modulation, defined as follows \cite{farace}:
\begin{equation}\label{e_n_period}
 E_N=\underset{t\in\left[\mathcal{T},\mathcal{T}+\tau\right]}{\mathrm{max}}E_N(t),
\end{equation}
after a long enough time $\mathcal{T}\gg1/\kappa,1/\gamma_m$. It is worth mentioning that, with the application of periodically modulated mechanical coupling, $\lambda_0/\omega_m=0.005$ we obtain a remarkable degree of stationary entanglement, $E_N=0.34$, as opposed to $E_N=0.04$ , which is achieved with no direct mechanical coupling.

\begin {figure}[t]\label{temp}
\begin {center}
\includegraphics [width =5cm]{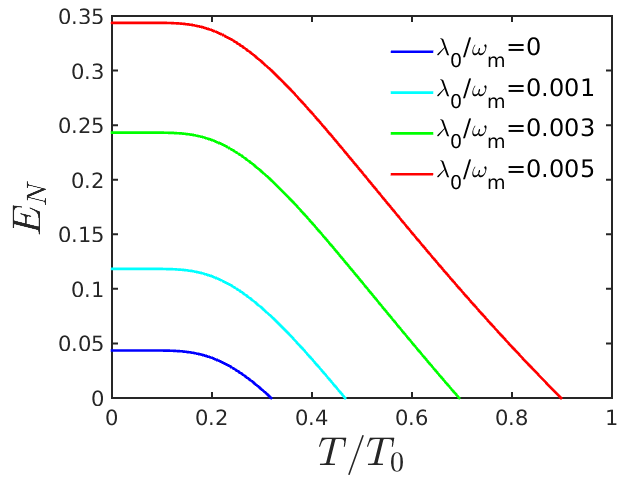}
\caption{\label {temp}(Color online) Dependence of stationary mechanical entanglement on the oscillator temperature. The other parameters are fixed as Fig. 2.}
\end{center}
\end{figure}
The dependence of the stationary mechanical entanglement on the oscillator temperature is exhibited in Fig. 3. As expected, it could be seen that the degree of entanglement decreases monotonically with increase in temperature $T$. However, one should note the improved robustness for the same mechanical entanglement with an increase in the coupling strength $\lambda_0$. For example, it is shown that in presence of the mechanical coupling $\lambda_0/\omega_m=0.005$, the degree of entanglement survives up to a relatively high temperature $T/T_0\approx0.9$, in sharp contrast to the case where entanglement is found to persist upto temperature $T/T_0\approx0.3$ in absence of the mechanical coupling.

Next, in Fig. 4(a), we once again depict the time evolution of $E_N$, similar to Fig. 2 but with a set of higher values of $\lambda_0/\omega_m$. Now one could observe that, with increase in $\lambda_0$, the entanglement not only grows much faster in time but also decays quickly to zero. Thus, it is evident that depending on the strength of the mechanical coupling, one could achieve a completely different dynamical behaviour of the mechanical entanglement. In order to investigate the role of oscillator temperature on the entanglement dynamics, we have redone the calculations for $T=3T_0$ and depicted it in Fig. 4(b). It exhibits that with the increase in temperature, $E_N$ decreases and there is a delay in the entanglement formation as well as reduction in survival time for entanglement, compared with its $T=0$ counterpart. It should be noted that this feature is also reported in Ref. \cite{roque}. However, if one compares Fig.(3) and Fig.4(b), it is clear that a significant degree of entanglement could be attained at a relatively high temperature, with higher mechanical coupling strength.

\begin {figure}[t]\label{en_un}
\begin {center}
\includegraphics [width =8.6cm, height=4cm]{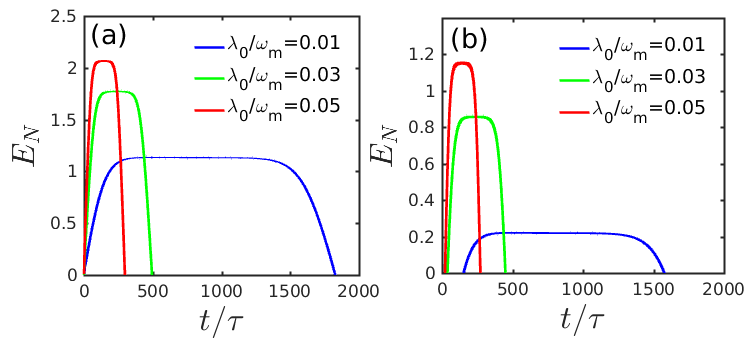}
\caption{\label {en_unstable}(Color online) Entanglement dynamics of the two coupled mechanical oscillators for $\lambda_0/\omega_m=[0.01,0.05]$ at (a) $T/T_0=0$ and (b) $T/T_0=3$. The other parameters are fixed as Fig. 2.}
\end{center}
\end{figure}

Now, to further probe into the entanglement dynamics and the role of the mechanical coupling on it, we introduce the normal modes for the mechanical oscillators as follows:
$\delta q_{\pm}=\left(\delta q_1 \pm \delta q_2\right)/\sqrt{2}$, $\delta p_{\pm}=\left(\delta p_1 \pm \delta p_2\right)/\sqrt{2}$, and, rewrite the linearized optomechanical Hamiltonian in the following way:
\begin{equation}\label{pm}
 H^\mathrm{lin}=H_++H_-,
\end{equation}
where, $H_{\pm}$ is given by
\begin{subequations}
\begin{align}
 H_+ &=\frac{\Delta}{2}\left(\delta X^2 + \delta Y^2 \right)+\frac{1}{2}\left(\omega_m \delta p_+^2 + \left(\omega_m+\lambda(t)\right)\delta q_+^2\right)\\\nonumber &-\sqrt{2}\left(G_x \delta X + G_y \delta G_y\right)\delta q_+,\\
 H_- &=\frac{1}{2}\left(\omega_m \delta p_-^2 + \left(\omega_m-\lambda(t)\right)\delta q_-^2\right).\label{m}
\end{align}
\end{subequations}
The above Hamiltonian \eqref{pm} describes two independent parametric oscillators, one of which ($+$ mode) is coupled to the cavity field via the usual optomechanical interaction, while the other ($-$ mode) one is completely free. Following a similar procedure, used to obtain Eq. \eqref{eqm_corr}, we construct the correlation matrix corresponding to the normal modes and the cavity field.

In order to illustrate the dynamics of the normal modes in the so-called phase-space, in Fig. 5 and 6 we depict the respective Wigner functions (Eq. \eqref{wigner})  at some specific times, for two different mechanical coupling strengths $\lambda_0/\omega_m=0.005$ and $\lambda_0/\omega_m=0.01$. We note that the value of the coupling strengths are so chosen to yield two different kinds of entanglement dynamics. The localization of the Wigner function in phase space, for $+$ mode, could be clearly seen in Fig. 5(a)-(c) and 6(a)-(c), and, this localization is independent of the mechanical coupling strength. On the contrary, the $-$ mode exhibits  both the localization and delocalization phenomena depending on the strength of the mechanical coupling. Fig. 5(d)-(f) shows localization of the $-$ mode when we have, $\lambda_0/\omega_m=0.005$, which is maintained even for a sufficiently long time $t/\tau=5000$. However, as could be seen from Fig. 6(d)-(f), with increase in $\lambda_0/\omega_m$, the delocalization occurs quickly in time: The Wigner function stretches along the dynamical rotating axis along with a contraction in the perpendicular direction. This feature become quite prominent if one observe the dynamics at a time $t/\tau=300.45$. This clearly signifies a dynamical instability corresponding to the $-$ mode. Therefore, we can infer that the asymptotic nature of the mechanical entanglement is directly related to the instability in the $-$ mode.
\begin {figure}[t]
\begin {center}
\includegraphics [width =8.2cm, height=6cm]{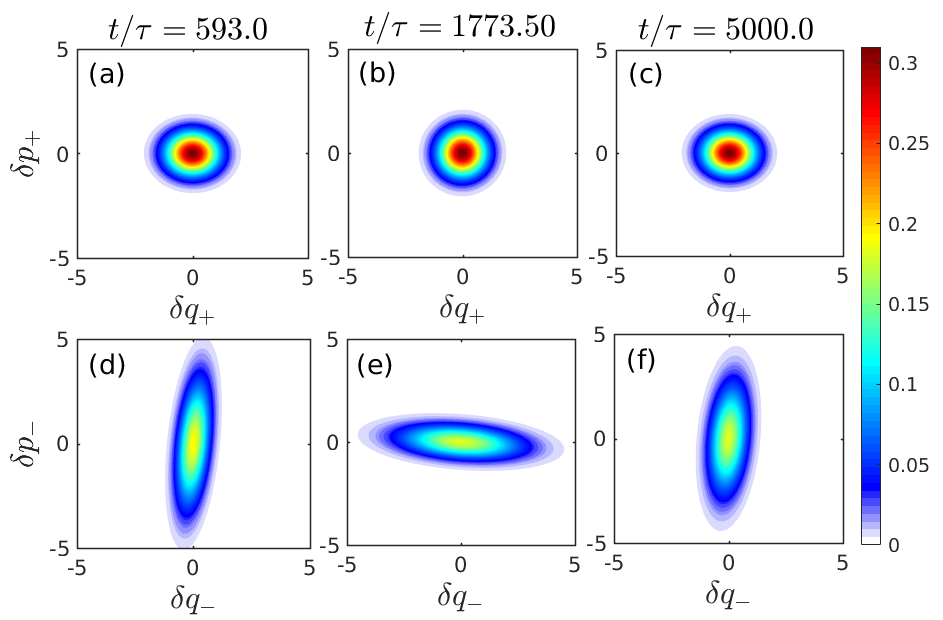}
\caption{\label {stability}(Color online) The wigner function for $+$ and $-$ mode at differetnt times, for the mechanical coupling stregth $\lambda_0/\omega_m=0.005$. The other parameters are fixed as Fig. 2.}
\end{center}
\end{figure}

Now, to explicitly derive the relationship between the entanglement dynamics and the strength of the mechanical coupling, we focus on the stability of the $-$ mode. Starting form the Hamiltonian \eqref{m}, we derive the equation of motion corresponding to the $-$ mode, as given below:
\begin{equation}\label{m_mode}
\ddot{\delta q}_-+\left(\omega_m^2-\omega_m\lambda_0 \cos\Omega t\right)\delta q_-+\gamma_m\dot{\delta q}_-=0.
\end{equation}
The above equation corresponds to a dissipative classical parametric oscillator, with a time modulated mechanical frequency $\omega_m^2(t)=\omega_m^2-\omega_m\lambda_0\cos(\Omega t)$. Following a substitution $\tilde{t}=\frac{\Omega t}{2}$, we can rewrite the Eq. \eqref{m_mode}, in the following form:
\begin{equation}\label{adim}
\ddot{\delta q}_-+\left(\tilde{\omega}_m-2\tilde{\lambda}_0 \cos 2\tilde{t}\right)\delta q_-+\tilde{\gamma}_m\dot{\delta q}_-=0,
\end{equation}
where the dimensionless parameters are defined as follows:
\begin{gather}
\tilde{\omega}_m=\frac{4\omega_m^2}{\Omega^2}, \ \tilde{\lambda}_0=\frac{2\omega_m\lambda_0}{\Omega_2}, \ \tilde{\gamma}_m=\frac{2\gamma_m}{\Omega}.
\end{gather}
Now, defining $\delta q_-=\tilde{\delta q}_-e^{-\tilde{\gamma}_m\tilde{t}/2}$ and substituting in Eq. \eqref{adim}, we get the standard form of the canonical Mathieu equation
\begin{equation}\label{final_mathieu}
\ddot{\tilde{\delta q}}_-+\left(\delta-2\epsilon\cos(2\tilde{t})\right)\tilde{\delta q}_-=0,
\end{equation}
where $\delta$ and $\epsilon$ are respectively given by $\delta=\frac{4\omega_m^2-\gamma_m^2}{\Omega^2}$ and $\epsilon=\frac{2\omega_m\lambda_0}{\Omega_2}$. It is clear that, for a modulation frequency $\Omega^2\approx4\omega_m^2$, we have $\delta\approx1$ and $\epsilon\ll1$. In this limit, one can neglect all the higher-order terms in the eigenvalues of the Mathieu's equation Eq.  \eqref{final_mathieu} (see appendix \ref{mathieu_equation} for a better discussion) and obtain:
\begin{subequations}\label{Mathieu_sol}
\begin{gather}
\alpha_1(\epsilon)\approx1+\epsilon,\\
\beta_1(\epsilon)\approx1-\epsilon.
\end{gather}
\end{subequations}
\begin {figure}[t]
\begin {center}
\includegraphics [width =8.2cm, height=6cm]{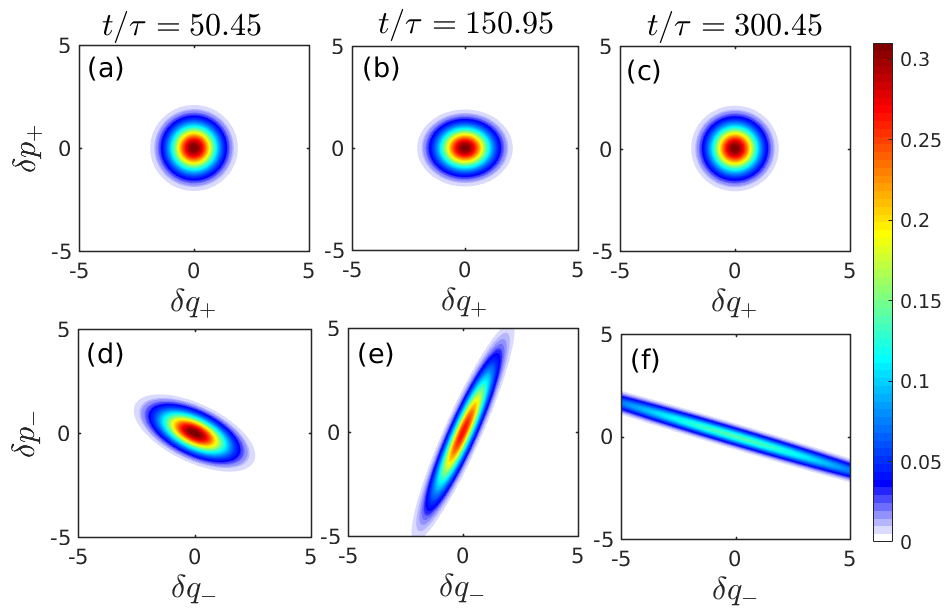}
\caption{\label {stability}(Color online) The wigner function for $+$ and $-$ mode at differetnt times, for the mechanical coupling strength $\lambda_0/\omega_m=0.01$. The other parameters are fixed as Fig. 2.}
\end{center}
\end{figure}

The stability of the $-$ mode in the $\epsilon$-$\delta$ plane, is depicted in Fig. 7. One can observe that, for the $-$ mode to be stable the following stability criteria must be satisfied:
\begin{equation}\label{stability_mathieu}
\beta_1(\epsilon)=1-\epsilon\leq\delta\leq\alpha_1(\epsilon)=1+\epsilon.
\end{equation}
Solving Eq. \eqref{stability_mathieu} in terms of the $\Omega$, $\omega_m$ and $\gamma_m$, we can obtain an analytical expression for the critical mechanical coupling strength $\lambda_{0c}$, given as follows:
\begin{equation}\label{crit}
\frac{\lambda_0}{\omega_m}\leq\frac{\lambda_{0c}}{\omega_m}=\frac{\Omega^2-4\omega_m^2+\gamma_m^2}{2\omega_m^2}.
\end{equation}
It should be noted that for the modulation frequency $\Omega_m=2.003\omega_m$, one gets the following critical mechanical coupling strength $\lambda_{0c}/\omega_m\approx0.0063$. This situation is further illustrated in Fig. 7, respectively, for the three distinct mechanical coupling strengths $\lambda_0/\omega_m=0.005$ (green circle), $\lambda_0/\omega_m=0.006$ (blue diamond), and $\lambda_0/\omega_m=0.007$ (red square). We can see that the points corresponding to the aforementioned coupling strengths, respectively, locates in the stable, on the boundary and in the unstable zone of the $-$ mode. This well justifies our previously obtained entanglement dynamics, corresponding to the different sets of the mechanical coupling strengths.
\begin {figure}[t]
\begin {center}
\includegraphics [width =5cm, height=4cm]{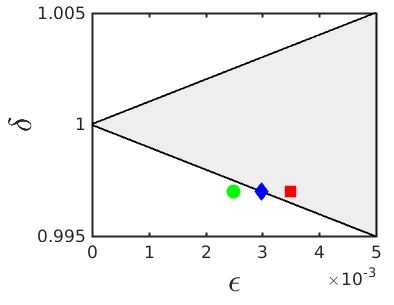}
\caption{\label {stability}(Color online) Stable (white) and unstable (grey) phase of the $-$ mode, for $\epsilon\ll1$. Here, the circle (green), diamond (blue) and square (red) respectively corresponds to the $\lambda_0/\omega_m=0.005, 0.006$ and $0.007$. The other parameters are fixed as Fig. 2.}
\end{center}
\end{figure}

Finally, in Fig. 8 we depict the dependence of the stationary mechanical entanglement on the modulation frequency. It can be seen that the entanglement $E_N$ is quite sensitive to the variation in the modulation frequency, which can be attributed directly to the instability in the $-$ mode. Furthermore, one can observe that the peak of the stationary entanglement is obtained exactly  at $\Omega/\omega_m=2.003$, which justifies our initial choice of the modulation frequency.

\section{CONCLUSION}\label{conclusion}

In conclusion, we have proposed a scheme to entangle two directly coupled mechanical oscillators, in an optomechanical system. Our scheme exploits the periodic modulation technique, in both the external driving and mechanical coupling strengths. We observe that an abrupt transition from stationary to dynamical mechanical entanglement occurs when the $-$ mode becomes unstable. More importantly, it is shown that in the presence of the mechanical coupling, a significant improvement in the robustness of the generated entanglement could be achieved with respect to the oscillators temperature. Finally, based on the  eigenvalues of the Mathieu's equations, we give an analytical estimations corresponding to the  the critical mechanical coupling strength, where the transition occurs. The feasibility of the chosen parameters, makes our proposed scheme a promising mean to realize macroscopic quantum entanglement within current state-of-the-art experimental setups.

\section*{ACKNOWLEDGEMENT}

S. Chakraborty would like to thank MHRD, Government of India for providing a financial support for his research.

\setcounter{section}{0}
\section*{appendix}
\subsection{\label{gaussian}Entanglement in Gaussian States}
\setcounter{equation}{0}
\renewcommand{\theequation}{A.\arabic{equation}}

In the context of continuous-variable (CV) quantum information, Gaussian states are of central importance. These are the states with Gaussian Wigner function and are completely characterized by its first and second moment of the field quadrature operators. For any $N$ mode Gaussian state, the vector of the first moments reads $\bar{R}=(\langle R_1 \rangle, \langle R_1 \rangle,...,\langle R_N \rangle, \langle R_N \rangle)$, while the second moments is denoted by the $2N\times2N$ covariance matrix (CM) $V$ of elements
\begin{equation}
 V_{ij}=\frac{1}{2}\left(\langle R_i R_j+ R_j R_i\rangle\right)-\langle R_i \rangle \langle R_j \rangle.
\end{equation}
\begin {figure}[t]
\begin {center}
\includegraphics [width =5.3cm, height=4cm]{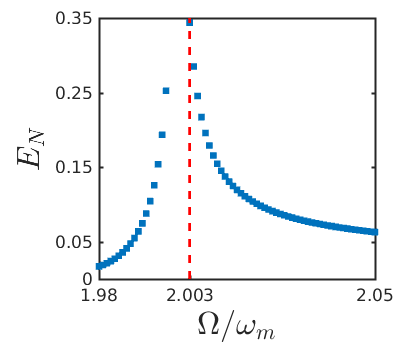}
\caption{\label {stability}(Color online) Dependence of stationary mechanical entanglement $E_N$ on the modulation frequency. The other parameters are fixed as Fig. 2.}
\end{center}
\end{figure}
Here, $R$ stands for the $2N$ dimensional vector of CV operators $R^T=(q_1,p_1,q_2,p_2,,...,q_N,p_N)$. However, by following a local unitary transformation the first moments could be easily adjusted to zero, without affecting any informationally relevant properties. With this consideration, the Wigner function for a $N$ mode Gaussian state could be written as follows \cite{conti}
\begin{equation}\label{wigner}
 W(R)=\frac{1}{(2\pi)^N\sqrt{\mathrm{Det}V}}e^{-\frac{1}{2}R^TV^{-1}R}.
\end{equation}

Now, to discuss entanglement in CV systems, we consider a very prototypical CV entangled state, \textit{i.e.} a two-mode Gaussian state. This type of state can be represented by the following covariance matrix
\begin{align}\label{eq:14}
V_2=\left(
  \begin{array}{cc}
    A & C \\
    C^T & B \\
  \end{array}
\right),
\end{align}
where $A$, $B$ and $C$ are $2\times2$ block matrices, respectively,  describing the local properties mode $A$, mode $B$ and the intermode correlation between $A$ and $B$. The degree of entanglement between the two modes is calculated by the so-called logarithmic negativity $E_N$ \cite{vidal, adesso},  defined as:
\begin{align}\label{eq:16}
E_N=\mathrm{max}\left[0,-\ln2\nu^-\right].
\end{align}
Here $\nu^-\equiv2^{-1/2}\left[\Sigma(V_2)-\sqrt{\Sigma(V_2)^2-4\mathrm{det}V_2}\right]^{1/2}$ is the smallest symplectic eigenvalue of the partial transpose of $V_2$ with $\Sigma(V_2)\equiv \mathrm{det}(A)+\mathrm{det}(B)-2\mathrm{det}(C)$. A Gaussian state is said to be entangled ($E_N>0$) if and only if $\nu^-<1/2$ which is equivalent to Simon's necessary and sufficient nonpositive partial transpose criteria \cite{simon}.

\subsection{\label{mathieu_equation}Matheu's Equation}
\setcounter{equation}{0}
\renewcommand{\theequation}{B.\arabic{equation}}

The canonical form of Matheu's equation for the parameters $\delta$ and $\epsilon$ is given by \cite{mathieu1, mathieu2}
\begin{equation}\label{can_mathieu}
 \ddot{y}+\left(\delta-2\epsilon \cos(2t)\right)y=0.
\end{equation}
This equation is a linear second order differential equation, with periodic coefficients. In general, the solution of such equation varies depending on the choice of $\delta$ and $\epsilon$. However, it should be noted that to maintain the periodicity of the solution, $\delta$ and $\epsilon$ must be interrelated. Therefore, one has a set of eigenvalues $\alpha_n(\epsilon)$ $(n=0,1,2,3,...)$ and $\beta_n(\epsilon)$ $(n=0,1,2,3,...)$ necessary to yield solution of Eq. \eqref{can_mathieu}. For small $\epsilon$, the first three $\alpha_n(\epsilon)$ and $\beta_n(\epsilon)$ could be written as follows \cite{mathieu1}
\begin{subequations}
 \begin{align}
  \alpha_o &=-\frac{1}{2}\epsilon^2+\frac{7}{128}\epsilon^4+O(\epsilon^6), \\
  \alpha_1 &=1+\epsilon-\frac{1}{8}\epsilon^2-\frac{1}{64}\epsilon^3+O(\epsilon^4),\\
  \beta_1 &=1-\epsilon-\frac{1}{8}\epsilon^2+\frac{1}{64}\epsilon^3+O(\epsilon^4).
 \end{align}
\end{subequations}

\end{document}